\documentclass[prd,twocolumn,showpacs,eqsecnum,amsfonts,amssymb,
amsmath,a4paper,english,superscriptaddress
]{revtex4}
\usepackage{graphicx,multirow}
\usepackage[innercaption]{sidecap}
\newlength{\inda}
\newlength{\indb}

\def\g{\text{\sl g}}

\def\x{{\mathbf x}}

\def\im{{\rm i}}

\def\det{{\mathrm{det}}}

\def\Box{\kern0.5pt{\lower0.1pt\vbox{\hrule height.5pt width 6.8pt
  \hbox{\vrule width.5pt height6pt \kern6pt \vrule width.3pt}
  \hrule height.3pt width 6.8pt} }\kern1.5pt}

\begin{document}
\setcounter{topnumber}{1}

\title{Stability analysis of sonic horizons in Bose-Einstein condensates}
\author{C. Barcel\'{o}}
\affiliation{Instituto de Astrof\'{i}sica de Andaluc\'{i}a, CSIC, Camino Bajo de Hu\'{e}tor 50, 18008 Granada, Spain}
\author{A. Cano}
\affiliation{Instituto de Astrof\'{i}sica de Andaluc\'{i}a, CSIC, Camino Bajo de Hu\'{e}tor 50, 18008 Granada, Spain}
\author{L. J. Garay}
\affiliation{Departamento de F\'{i}sica Te\'{o}rica II, Universidad
Complutense de Madrid, 28040 Madrid, Spain}
\affiliation{Instituto de Estructura de la Materia, CSIC, Serrano 121,
28006 Madrid, Spain}
\author{G. Jannes}
\affiliation{Instituto de Astrof\'{i}sica de Andaluc\'{i}a, CSIC, 
Camino Bajo de Hu\'{e}tor 50, 18008 Granada, Spain}
\affiliation{Instituto de Estructura de la Materia, CSIC, Serrano 121,
28006 Madrid, Spain}

\date{\today}

\begin{abstract}
We examine the linear stability of various configurations in
Bose-Einstein condensates with sonic horizons. These configurations
are chosen in analogy with gravitational systems with a black hole
horizon, a white hole horizon and a combination of both. We discuss
the role of different boundary conditions in this stability analysis,
paying special attention to their meaning in gravitational terms. We
highlight that the stability of a given configuration, not only
depends on its specific geometry, but especially on these boundary
conditions. Under boundary conditions directly extrapolated from those
in standard General Relativity, black hole configurations, white hole
configurations and the combination of both into a black hole--white
hole configuration are shown to be stable. However, we show that under
other (less stringent) boundary conditions, configurations with a
single black hole horizon remain stable, whereas white hole and black
hole--white hole configurations develop instabilities associated to
the presence of the sonic horizons.
\end{abstract}

 \maketitle

\section{Introduction}
\label{S:introduction}

It is widely expected that underneath the general relativity description
of gravitational phenomena there is a deeper layer
in which quantum physics plays an important role. However, at this
stage we don't have enough intertwined theoretical and observational
knowledge to know how an appropriate description of what underlies
gravity is or should be. Moreover, starting from
structurally-complete quantum theories of gravity, it could still be
very difficult to extract the specific way in which the first
``quantum'' modifications to classical general relativity might show
up (this happens for example within the Loop Quantum Gravity
approach~\cite{LQG}).

Analogue models of General Relativity
(GR)~\cite{analogue-book,review} provide specific and clear examples
in which effective spacetime structures ultimately emerge from
(non-relativistic) quantum many-body systems. For certain
(semiclassical) configurations and low levels of resolution, one can
appropriately describe the physical behaviour of the system by means
of a classical (or quantum) field theory in a curved (Lorentzian)
background geometry. However, when one probes the system with higher
and higher resolution, the geometrical structure progressively
dissolves into a purely quantum regime \cite{barcelo-bec}.
Therefore, although analogue models cannot be considered at this
stage complete models of quantum gravity (they do not lead to the
Einstein equations in any regime or approximation), they provide
specific and tractable models that reproduce many aspects of the
overall scenario expected in the realm of real gravity.

The main objective of this and similar studies is to obtain specific
indications about the type of deviations from the GR behaviour to be
expected when quantum gravitational effects become important. All,
under the assumption that the underlying structure to GR is somewhat
similar to that in condensed matter systems.
In particular, in this paper we are interested in the behaviour of
gravity-like configurations containing horizons within
Bose--Einstein condensates (BECs). (See, e.g., Refs.
\mbox{\cite{dalfovo,castin,ketterle}} and
\cite{barcelo-bec,sonicBH1,sonicBH2} for reviews on BECs and for
their usefulness as analogue models respectively). A nice feature of
these systems is that their theoretical description in terms of the
Gross-Pitaevskii (GP) equation can be interpreted as incorporating,
from the start, the first ``quantum'' corrections to the behaviour
of the system. Linear perturbations over a background BEC
configuration satisfy an equation which is a standard wave equation
over a curved effective spacetime plus corrections containing
$\hbar$. These corrections cause the dispersion relations in BEC to
be ``superluminal'' (strictly speaking, supersonic): some
perturbations can travel faster than the speed of sound in the
system. The effects of these corrections in the linearized dynamical
evolution of a configuration are especially relevant in the presence
of horizons as their one-way-membrane nature simply disappears. This
is in tune with the idea that a horizon can serve as a magnifying
glass of the physics at high energies (see, e.g.,
Ref.~\cite{Padmanabhan:1998vr}).

The specific objective of this paper is to analyze the dynamical
behaviour of (effectively) one-dimensional BECs with density and
velocity profiles containing one or two sonic horizons. In
particular, we search for the presence of dynamical instabilities
and analyze how their existence is related to the occurrence of
these horizons. The stability analysis presented in Ref.
\cite{sonicBH2} for black hole-like configurations with fluid sinks
in their interior, concluded that these configurations were
intrinsically unstable. However, the WKB analysis of the stability
of horizons in Ref.~\cite{leonhardt} suggested that black hole
horizons might well be stable, while configurations with white hole
horizons seem to posses unstable modes. Regarding configurations in
which a black hole horizon is connected with a white hole horizon in
a straight line, the analysis presented in Ref.~\cite{jacobson}
concluded, also within a WKB approximation, that these
configurations were intrinsically unstable, producing a so-called
``black hole laser''. However, when the white hole horizon is
connected back to the black hole horizon to produce a ring, it was
found that these configurations can be stable or unstable depending
on their specific form \cite{sonicBH1,sonicBH2}. This suggests that
periodic boundary conditions can eliminate some of the instabilities
associated with the black hole laser.

In this paper we will try to shed some light on all of these issues
and clear up some of the apparent contradictions. To simplify
matters, we will consider one-dimensional profiles that are
piecewise uniform with either one or two step-like discontinuities.
As has been discussed in Ref.~\cite{sonicBH2}, in terms of dynamical
(in)stability, there seems to be no crucial qualitative difference
between the present case and a profile with smooth transitions
between regions with an (asymptotically) uniform density
distribution. Therefore the idealized case that we consider here
should contain all the essential information relevant to more
complicated profiles as well. The specific way to examine the kind
of instability we are interested in, consists basically in seeking
whether, under appropriate boundary conditions, there are complex
eigenfrequencies of the system which lead to an exponential increase
with time of the associated perturbations, i.e. a dynamical
instability. Throughout the paper we will use a language and
notation as close to GR as possible. In particular, we will use
boundary conditions similar to those imposed in the standard
quasi-normal mode analysis of black holes in
GR~\cite{quasi-normal-modes}. One of the main results of our
analysis is to highlight the fundamental importance of the boundary
conditions in determining whether a configuration is stable or
unstable.

The structure of this paper is as follows. In the next section we
will review the basic ingredients of gravitational analogies in
BECs. At the same time, we will set up the conceptual framework for
our discussion, based on a parametrization well adapted for an
acoustic interpretation (a brief comparison with the Bogoliubov
representation is presented in appendix~\ref{A:app}).
Section~\ref{S:dynamical} contains a detailed formulation of our
specific problem. This includes the mode expansion in uniform
sections, a derivation of the matching conditions at each
discontinuity and a discussion of the various boundary conditions to
be applied. Then, in section~\ref{S:case-by-case} we proceed case by
case, analyzing different situations and presenting the results we
have obtained for each of them. This includes a brief description of
the numerical algorithm we have used. Finally, in
section~\ref{S:discussion} we discuss our results, compare them with
other results available in the literature, and draw some
conclusions.

\section{Preliminaries}
\label{S:preliminaries}

In second quantization, a dilute gas of interacting
bosons can be described by a quantum field $\widehat \Psi$ satisfying the
equation
\begin{eqnarray}
 \im \hbar \; \frac{\partial }{\partial t} {\widehat \Psi}=
 \left( - {\hbar^2 \over 2m} \nabla^2 +
 V_{\rm ext}(\x)
 +\g{\widehat \Psi}^{\dagger}{\widehat \Psi} \right){\widehat \Psi},
\end{eqnarray}
where $m$ is the boson mass, $V_\text{ext}$ the external potential
and $\text{\sl g}$ the coupling constant which is related to the
corresponding scattering length $a$ through $\g ={4\pi \hbar^2 a
/m}$. In this manner all quantum effects can, in principle, be taken
into account. Once the Bose-Einstein condensation has taken place,
the quantum field can be separated into a macroscopic wave function
$\psi$ (the corresponding order parameter) and a field operator
$\widehat \varphi$ describing quantum fluctuations over it:
${\widehat \Psi}=\psi+{\widehat \varphi}$. The macroscopic wave
function satisfies the Gross-Pitaevskii (GP) equation
\begin{align}
 \im \hbar \; \frac{\partial }{\partial t} \psi(t,\x)= \left(
 - {\hbar^2 \over 2m} \nabla^2
 + V_{\rm ext}(\x)
 + \g \; |\psi|^2 \right) \psi(t,\x),
\label{GP}
\end{align}
while for the linear quantum perturbation we have the Bogoliubov
equation
\begin{align}
 \im \hbar \; \frac{\partial }{\partial t} {\widehat \varphi}=
 \left( - {\hbar^2 \over 2m} \nabla^2 +
 V_{\rm ext}(\x)
 +\g \;2 |\psi|^2 \right){\widehat \varphi} +
 \g \; \psi^2 \; {\widehat \varphi}^{\dagger}.
 \label{bogolubov}
\end{align}
Adopting the Madelung representation for the order parameter
\begin{eqnarray}
\psi = \sqrt{n}e^{\im \theta/\hbar} e^{- \im \mu t/\hbar }
\label{madelung}
\end{eqnarray}
(here $n$ is the condensate density, $\mu$ the chemical
potential and $\theta$ a phase factor which is related to the
velocity potential), and substituting in (\ref{GP}) we arrive
at
\begin{subequations}
\begin{align}
\partial _t n &= - {1 \over m} \nabla \cdot (n \nabla \theta), \label{GP_n_theta_a}\\
\partial _t \theta &= -{1\over 2 m}(\nabla \theta)^2
- \g \; n
-V_\text{ext}-\mu -V_\text{quantum},
\label{GP_n_theta_b}\end{align}
\label{GP_n_theta}
\end{subequations}
where the so-called ``quantum potential'' is defined as
\begin{eqnarray}
V_\text{quantum}=
-{\hbar^2 \over 2 m}{\nabla^2 \sqrt{n}\over \sqrt{n}}.
\label{QP}
\end{eqnarray}
In most situations the quantum potential in Eq. \eqref{GP_n_theta_b}
can be neglected (see below). The resulting equations
\eqref{GP_n_theta_a} and \eqref{GP_n_theta_b} are then equivalent to
the continuity equation and the Bernoulli equation for a classical
fluid. In this case, it is well known that the propagation of acoustic
waves in the system can be described by means of an effective metric,
thus providing the analogy with the propagation of fields in curved
spacetimes~\cite{Unruh:1980cg,visser97}. Given a background
configuration ($n_0$ and $\theta_0$), this metric can be written as
\begin{align}
(g_{\mu \nu})=\frac{m}{\g} c
\begin{pmatrix}
v^2-c^2 && -\mathbf v^\text{T}\\
-\mathbf v && \mathbf \openone
\end{pmatrix},
\label{metric}
\end{align}
where $c^2 \equiv \g n_0/m$ and $\mathbf v \equiv \nabla \theta_0 /m$.
These magnitudes, $c$ and $\mathbf v$, represent the local velocity of
sound and the local velocity of the fluid flow respectively.

The functions $c(t,\x)$ and $\mathbf v(t,\x)$ completely
characterize the acoustic metric. In GR any metric has to be
obtained by solving the Einstein equations. Here, however, the
magnitudes $c(t,\x)$ and $\mathbf v(t,\x)$, and so the acoustic
metric, are those satisfying the continuity and Bernoulli equations
of hydrodynamics [Eqs. (\ref{GP_n_theta}) without the quantum
potential]. Thus, these equations play a role analogous to the
vacuum Einstein equations in GR. Of course, at the global non-linear
level these equations are completely different from the real
Einstein equations. But their way of acting when linearized
around a background solution captures the essence of a proper
linearized GR behaviour.

There exist, however, situations in which the quantum potential
in Eq. (\ref{GP_n_theta}) cannot be neglected. This is evidently the
case if the characteristic length of the spatial variations of the
condensate density is much smaller than the so-called healing length:
$\xi \equiv \hbar / (mc)$. But this case is not the only one. To
illustrate this point, let us consider the dispersion law obtained for
a homogeneous BEC (see below):
\begin{align}
(\omega - vk)^2= c^2k^2 + {1 \over 4} c^2 \xi^2 k^4.
\label{dispersion}
\end{align}
This is a ``superluminally modified'' dispersion relation due
to the presence of the term with $k^4$. For $\xi k \ll 1$ we
can rewrite this expression as
\begin{equation}
\omega = \left ( v \pm c\sqrt{1+ {1 \over 4}\xi^2k^2} \right )k \simeq
(v \pm c)k + {1 \over 8} c\xi^2 k^3 + \mathcal{O}(\xi^4k^5).
\end{equation}
Here we clearly see that the (relative) importance of the term
$\propto k^3$, given by ${c \over 8(v \pm c)}\xi^2 k^2$, depends not
only on the ratio between the corresponding wavelength and the
healing length, $\xi k$, but also on the specific features of the
background magnitudes [note that the factor $c/(v \pm c)$ may be
quite large]. In other words, the smallness of the corresponding
wavelength is a necessary condition in order to neglect the
contribution of the quantum potential, but it is not a sufficient condition. One has to bear this issue in mind, especially when the
system possesses horizons (i.e. points at which $c^2 = v^2$).

Summarizing, there are background configurations which, when probed
with sufficiently large wavelengths, act as if they were effective
Lorentzian geometries. But there are other configurations for which
this geometrical interpretation fails, irrespective of the probing
wavelength. The latter situation occurs when there are horizons in
the configuration: strictly speaking, we cannot talk about an
``effective Lorentzian geometry'' in the regions surrounding these
horizons.

Without forgetting this subtlety, we will continue to call
(\ref{metric}) the ``effective metric'' in the system, even when
analyzing the full GP equation. Then, we can consider the equations
(\ref{GP_n_theta}) to play the role of some sort of semiclassical
vacuum Einstein equations. Their treatment is classical, but they
incorporate corrections containing $\hbar$. Therefore, BECs'
standard treatment based on the GP equation provides an example of a
way of incorporating quantum corrections to the dynamics of a system
without recurring to the standard procedures of back-reaction.
Again, although at the global non-linear level these equations bear
no relationship whatsoever with any sort of "semiclassical" Einstein
equations, at the linear level, that is, in terms of linear
tendencies of departing from a given configuration, equations
(\ref{GP_n_theta}) encode the essence of the \emph{linearized} GR
behaviour (a Lorentzian wave equation in a curved background),
semiclassically modified to incorporate a superluminal dispersion
term, as we have already discussed.

We will now proceed to describe the details of our
specific calculations.

\section{Dynamical analysis}
\label{S:dynamical}

As a first step in our calculations, let us linearize the Eqs.
(\ref{GP_n_theta}).
Let us write
\begin{subequations}
\begin{align}
n (\x, t) &= n_0(\x) + \g^{-1} \widetilde n_1(\x, t),
\\
\theta(\x, t) &= \theta_0(\x) + \theta_1(\x, t),
\end{align}
\end{subequations}
where $\widetilde n_1$ and $\theta_1$ are small perturbations of the
density and phase of the BEC. The Eqs. (\ref{GP_n_theta}) then
separate into two time-independent equations for the background,
\begin{subequations}\label{GP_background}
\begin{align}
0&= - \nabla \cdot (c^2 \mathbf v), \label{GP_BG1}\\
0 &= -{1\over 2 }m\mathbf v^2 -m c^2 - V_\text{ext}-\mu
+{\hbar^2 \over 2m}{\nabla^2 c\over c},
\end{align}
\end{subequations}
plus two time-dependent equations for the perturbations,
\begin{subequations}
\label{GP_lin}
\begin{align}
\partial _t \widetilde n_1 &= - \nabla \cdot
\left( \widetilde n_1 \mathbf v + c^2 \nabla \theta_1 \right),
\label{GP1_lin}
\\
\partial _t \theta_1
&= -\mathbf v \cdot \nabla \theta_1 - \widetilde n_1 +{1 \over 4} \xi^2
\nabla \cdot \left[c^2 \nabla \left( {\widetilde n_1 \over
c^2}\right)\right].
\label{GP2_lin}
\end{align}
\end{subequations}

We will restrict ourselves to work in (1+1) dimensions. This means
that we consider perturbations propagating in a condensate in such a
way that the transverse degrees of freedom are effectively frozen.
In other words, the only allowed motions of both the perturbations and the
condensate itself are along the $x$-axis.

We will examine two types of one-dimensional background
profiles. The first type consists of two regions each with a uniform
density and velocity, connected through a step-like discontinuity. The
second type of profiles consists of three homogeneous regions, and
hence two discontinuities. We wish to know whether these profiles do
or do not present dynamical instabilities. The underlying question is
the relation between the presence of horizons and these dynamical
instabilities.
At each discontinuity, matching conditions apply that connect the
magnitudes describing the condensate at both sides of the
discontinuity. Furthermore, we need a set of boundary
conditions, which determine what happens at the far ends of the
condensate. Finally, in the uniform sections of the condensate, the
regime can either be subsonic or supersonic, and of course there
will be an acoustic horizon at each transition between a
subsonic and a supersonic region. All these elements determine the
characteristics of the system, and hence its eigenfrequencies.

We will now describe these elements one by one in detail.

\subsection{Plane-wave expansion in uniform regions}
\label{S:mode-expansion}

In order to study the dynamics of the system, let us first consider a
region in which the condensate is homogeneous (with $c$ and $v$
constant), and seek for solutions of Eqs. \eqref{GP_lin} in the form
of plane waves:
\begin{subequations}\label{plane-waves}\begin{align}
\widetilde n_1(x,t)&=
A e^{i(k x - \omega) t}, \\
\theta_1(x,t)&= B e^{i(k x - \omega) t},
\end{align}\end{subequations}
where $A$ and $B$ are constant amplitude factors. Our aim is to
elucidate about the possible instabilities of the system, so the
frequency $\omega$ and the wavevector $k$ in these expressions will be
considered as complex hereafter [the existence of solutions with
Im$(\omega)>0$ would indicate the instability of the
system]. Substituting into Eqs. \eqref{GP_lin} we find
\begin{align}
\begin{pmatrix}
i(\omega- vk)  && c^2 k^2 \\ \\
1 +{1 \over 4} \xi^2 k^2 && - i(\omega-vk)
\end{pmatrix}
\begin{pmatrix}
A\\ \\
B\end{pmatrix}=0.
\end{align}
For a non-trivial solution to exist, the determinant of the above
matrix must vanish. This condition gives the dispersion law
\eqref{dispersion} and since this is a fourth order polynomial in $k$,
its roots will, in general, give four independent solutions for the
equations of motion in the form \eqref{plane-waves}.

\subsection{Matching conditions at a discontinuity}
\label{S:matching}

Let us take $x=0$ to be a point of discontinuity. The values of $v$
and $c$ both undergo a finite jump when crossing this point. 
These jumps have to satisfy the background constraint $vc^2 =
\text{const}$ [see Eq. \eqref{GP_BG1}].
The solutions of Eqs. \eqref{GP_lin} in the regions $x<0$ and $x>0$
have the form of plane waves which are then subject to matching
conditions at $x=0$. It is not difficult to see that $\theta_1$ has
to be continuous at the jump but with a discontinuous derivative,
while the function $\widetilde n_1$ has to undergo a finite jump.
The exact conditions can be obtained by integrating Eqs.
\eqref{GP_lin} about an infinitesimal interval containing the point
$x=0$. This results in the following four independent\emph{,
generally valid} matching conditions:
\begin{subequations}\label{matching}\begin{align}
[\theta_1]=0,\qquad [v \widetilde n_1+ c^2 \partial_x \theta_1]=0,\label{matching_a}\\
\left[{\widetilde n_1\over c^2}\right]=0,
\qquad \left[c^2\partial_x
\left({\widetilde n_1\over c^2}\right)\right]=0.
\end{align}\end{subequations}
The square brackets in these expressions denote, for instance,
$[\theta_1]=\left.\theta_1\right|_{x=0^+} -\left.
\theta_1\right|_{x=0^-}$. We can simplify the second condition in
Eq.\eqref{matching_a} to $[c^2 \partial_x \theta_1]=0$ by noting that
$[v \widetilde n_1]=0$ because of the background continuity equation
\eqref{GP_BG1}, while for our choice of a homogeneous background the
last condition becomes simply $[\partial_x \widetilde n_1]=0$.

For a given frequency $\omega$, the general solution of
Eqs. \eqref{GP_lin} can be written as
\begin{widetext}
\begin{align}
\widetilde n_1 = \begin{cases}
\displaystyle\sum _{j=1}^4 A_j e^{i(k_jx - \omega t)}& (x<0),\\
\displaystyle\sum _{j=5}^8 A_j e^{i(k_jx - \omega t)}& (x>0),
\end{cases}
\qquad \qquad \theta_1 =
\begin{cases}
\displaystyle\sum _{j=1}^4 A_j {\omega - v_\text{L}
k_j\over ic^2_\text{L} k ^2_j} e^{i(k_jx- \omega t)}  & (x<0),\\
\displaystyle\sum _{j=5}^8 A_j {\omega - v_\text{R} k_j\over
ic^2_\text{R} k ^2_j}e^{i(k_jx- \omega t)}& (x>0),
\end{cases}
\end{align}
where $\{ k_j\}$ are the roots of the corresponding dispersion
equations (four roots for each homogeneous region), and the constants $A_j$
have to be such that the matching conditions
(\ref{matching}) are satisfied. The subscripts $L$ and $R$
indicate the values of $c$ and $v$ in the left-hand-side (lhs) and
the right-hand-side (rhs) region respectively. We can write down
these conditions in matrix form $\Lambda_{ij}A_j=0$, where
\begin{eqnarray}
\left (\Lambda_{ij} \right )=
\begin{pmatrix}
{\omega-v_\text{L} k_1\over c^2_\text{L}k ^2_1} &
{\omega-v_\text{L} k_2\over c^2_\text{L}k ^2_2} &
{\omega-v_\text{L} k_3\over c^2_\text{L}k ^2_3} &
{\omega-v_\text{L} k_4\over c^2_\text{L}k ^2_4} &
-{\omega - v_\text{R} k_5\over c^2_\text{R}k ^2_5}&
-{\omega - v_\text{R} k_6\over c^2_\text{R}k ^2_6}&
-{\omega - v_\text{R} k_7\over c^2_\text{R}k ^2_7}&
-{\omega - v_\text{R} k_8\over c^2_\text{R}k ^2_8}
\\ \\
{\omega\over k_1}&{\omega\over k_2}&{\omega\over k_3}&{\omega\over k_4}&
- {\omega\over k_5}&- {\omega\over k_6}&- {\omega\over k_7}&- {\omega\over k_8}
\\ \\
{1\over c^2_\text{L}}&{1\over c^2_\text{L}}&{1\over c^2_\text{L}}&{1\over c^2_\text{L}}&
- {1\over c^2_\text{R}}&- {1\over c^2_\text{R}}&- {1\over c^2_\text{R}}&- {1\over c^2_\text{R}}&
\\ \\
k_1 &k_2 &k_3 &k_4 &- k_5&- k_6&- k_7&- k_8&
\\
\end{pmatrix}.
\label{lambda-matrix}
\end{eqnarray}
\end{widetext}
Furthermore, these conditions have to be complemented with
conditions at
the boundaries of the system and then we will obtain the
solution of a particular problem.

\subsection{Boundary conditions}
\label{S:boundary}

In order to extract the possible intrinsic instabilities of a BEC
configuration, we have to analyze whether there are linear mode
solutions with positive $\text{Im}(\omega)$ that satisfy outgoing
boundary conditions. By ``outgoing'' boundary conditions we mean that
the group velocity is directed outwards (toward the boundaries of the
system). The group velocity for a particular $k$-mode is defined as
\begin{equation}\label{group-velocity}
v_g \equiv \text{Re} \left( \frac{d \omega}{dk}\right) =
\text{Re}\left(
\frac{c^2 k + {1\over 2} \xi^2 c^2 k^3}{\omega - vk} + v
\right),
\end{equation}
where we have used the dispersion relation \eqref{dispersion}. The
physical idea behind this outgoing boundary condition is that only
disrupting disturbances originated inside the system can be called
instabilities.

To illustrate this assertion, let us look at the classical linear
stability analysis of a Schwarzschild black hole in GR. When
considering outgoing boundary conditions both at the horizon and in
the asymptotic region at infinity~\cite{quasi-normal-modes}, only
negative $\text{Im}(\omega)$ modes (the quasi-normal modes) are found,
and thus the black hole configuration is stable. If the presence of
ingoing waves at infinity were allowed, there would also exist
positive $\text{Im}(\omega)$ solutions. In other words, the
Schwarzschild solution in GR is stable when considering only internal
rearrangements of the configuration. If instead the black hole were
allowed to absorb more and more energy coming from infinity, its
configuration would continuously change and appear to be unstable.

The introduction of modified dispersion relations adds an
important difference with respect to the traditional boundary
conditions used in linearized stability analysis in GR. Consider for
example a black hole configuration. In
a BEC black hole, one boundary is the standard asymptotic region,
just like in GR. For an acoustic (quadratic)
dispersion relation, nothing can escape from the interior of a sonic
black hole (the acoustic behaviour is analogous to linearized GR). 
But due to the superluminal corrections,
information from the interior of the acoustic black hole can escape
through the horizon and affect its exterior. Therefore, since we are taking
this permeability of the horizon into consideration, the other
boundary is not the black hole horizon itself (usually described in
GR by an infinite value of the ``tortoise'' coordinate; see for
example~\cite{quasi-normal-modes}), but the internal singularity.
The outgoing boundary condition at such a singularity reflects the
fact that no information can escape from it.

There is another complication that deserves some attention. In the
case of a phononic dispersion relation, the signs of $v_g
\equiv v \pm c$ and $\text{Im}(k) \equiv (v \pm c)\text{Im}(\omega)$
coincide for $\text{Im}(\omega)>0$. For example, in an asymptotic \mbox{$x
\to +\infty$} subsonic region, an outgoing $k$-mode has $v_g>0$, so that
$\text{Im}(k)>0$ and, therefore, the mode is damped towards this
infinity (giving a finite contribution to its norm). Owing to this fact, in the
linear stability analysis of black hole configurations, it is usual to
assume that stable modes correspond to non-normalizable perturbations
(think of the standard quasi-normal modes), while unstable modes
correspond to normalizable perturbations.

When considering modified dispersion relations, however, this
association no longer holds. In particular, with the BEC
dispersion relation, in an asymptotic \mbox{$x \to +\infty$} region,
among the unstable ($\text{Im}(\omega)>0$) outgoing \mbox{($v_g>0$)}
$k$-modes, there are modes with $\text{Im}(k)>0$ as well as modes
with $\text{Im}(k)<0$. An appropriate
interpretation of these two possibilities seems to be the following. The
unstable outgoing modes that are convergent at infinity (those with
$\text{Im}(k)>0$) are associated with perturbations of the system
that are initiated in an internal compact region of the system.
Unstable outgoing modes that are divergent at infinity are
associated with initial perturbations acting also at the boundary at
infinity itself.

\begin{figure}[tbp]
\includegraphics[width=.4\textwidth,clip]{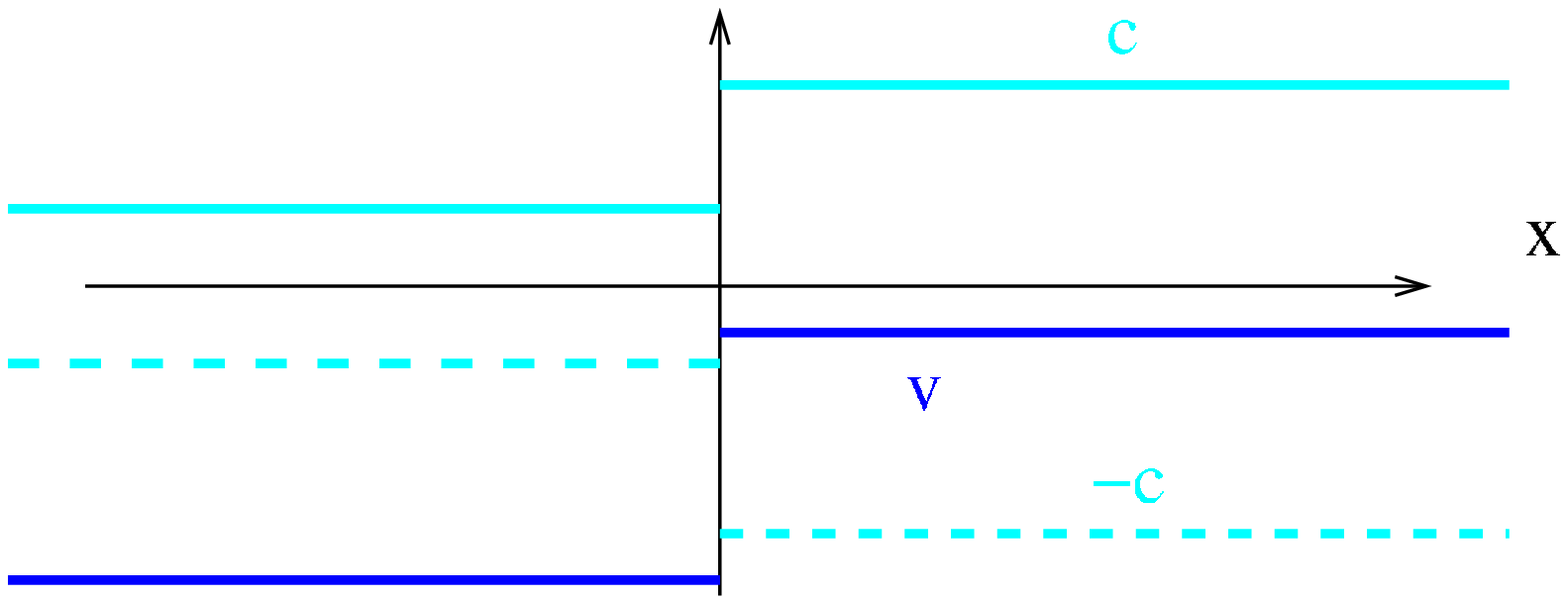}
\bigskip
\caption{Flow and sound velocity profiles with step-like discontinuities
simulating a black hole-like configuration. The negative
value of $v$ indicates that the fluid is left-moving. At the rhs, the
fluid is subsonic since $c>\lvert v \rvert$. At the lhs it has become
supersonic. At $x=0$, there is a sonic horizon.}
\label{F:v-super-sub}
\end{figure}

Take for example a black hole-like configuration of the form
described in Fig. \ref{F:v-super-sub}. The right asymptotic region,
which can be interpreted as containing a ``source'' of BEC gas in
our analogue model, simulates the asymptotic infinity outside the
black hole in GR. The convergence condition at the rhs then implies
that the perturbations are not allowed to affect this asymptotic
infinity initially. However, for the left asymptotic region, this
condition is less obvious. In our BEC configuration, this left
asymptotic region can be seen as representing a ``sink''. It
corresponds to the GR singularity of a gravitational black hole. The
fact that in GR this singularity is situated at a finite distance
(strictly speaking, at a finite amount of proper time) from the
horizon, indicates that it might be sensible to allow the
perturbations to affect this left asymptotic region from the start.
We will therefore consider two possibilities for the boundary
condition at $x \rightarrow -\infty$. (a) Either we impose
convergence in both asymptotic regions, thereby eliminating the
possibility that perturbations have an immediate initial effect on
the sink, or \mbox{(b) we} allow the perturbations to affect the sink right
from the start, i.e. we don't impose convergence at the left
asymptotic region. The option of imposing the convergence at the
left asymptotic region could be interpreted as excluding the
influence of the singularity on the stability of the system. In
other words, condition (b) would then be equivalent to examining the
stability due to the combined influence of the horizon and the
singularity, while under condition (a) only the stability of the
horizon would be taken into account.

As a final note to this discussion, since we are interested in
the analogy with gravity, we have assumed an infinite system at the
rhs. In a realistic condensate other boundary conditions could
apply, for example taking into account the reflection at the ends of
the condensate (see e.g. \mbox{\cite{sonicBH1,sonicBH2})}.

\section{Case by case analysis and results}
\label{S:case-by-case}

We will now briefly describe the general calculation method which we
have used, and then discuss case by case the specific
configurations we have analyzed.

\subsection{Numerical method}
\label{S:num}

We first consider background flows and sound velocity profiles with
one discontinuity. We will always assume left-moving flows.

We are seeking for possible solutions of the linearized Eqs.
(\ref{GP_lin}) with $\text{Im}(\omega)>0$. We use the following
numerical method:

\begin{enumerate}

\item For each frequency $\omega$ in a grid covering an
appropriate region of the upper-half complex plane, we calculate its
associated $k$-roots [by solving the dispersion relation
(\ref{dispersion})] and their respective group velocities
(\ref{group-velocity}) at both the lhs and the rhs of the
configuration.

\item
We then take the four equations $\Lambda_{ij}A_j=0$, where $\Lambda$
is the $4\times 8$ matrix (\ref{lambda-matrix}) determined by the
matching conditions at the discontinuity. For each mode $k_j$ that
does not satisfy the boundary conditions in the relevant asymptotic
region, we add an equation of the form $A_j=0$. Thus we have a total
set of equations which can be written as
$\widetilde\Lambda_{ij}A_j=0$, where $\widetilde\Lambda$ is now a
$(4+N) \times 8$ matrix the number $N$ of forbidden modes can in
principle vary between $0$ and $8$). Numerically it is convenient to
normalize $\widetilde\Lambda$ in such a way that its rows are unit
vectors.

\item We can then define a non-negative function $F(\omega)$, where
$F(\omega)=0$ means that the frequency $\omega$ is an eigenfrequency
of the system, in the following way.

\begin{itemize}

\item
If $N<4$, then $F(\omega) = 0$. Indeed, we have $8$ variables
$A_j$ and $4+N$ equations. Then, it is obvious that there will always
exist a non-trivial solution $\{A_j\}$.

\item
If $N=4$, then $F(\omega) = |\det(\widetilde\Lambda)|$. In this case
there will be a non-trivial solution only if the
determinant of the matrix $\widetilde\Lambda$ vanishes.

\item
If $N=5$, then $F(\omega) $ is taken to be the sum of the modulus of
all possible determinants that are obtained from $\widetilde \Lambda $ by
eliminating one row. Notice that in this case $\widetilde \Lambda $ is
a non-square \mbox{$9\times 8$} matrix because there are more conditions
than variables. In this situation it is highly unexpectable to find
zeros in $F$ as this would mean a double degeneracy.

\item
If $N>5$, $F(\omega) $ is defined by a straightforward generalization
of the procedure for $N=5$.

\end{itemize}

\item
We plot the function $F(\omega)$ in the upper half of the complex plane,
and look for its zeros. Each of these zeros indicates an unstable
eigenfrequency, and so the presence (or absence) of these zeros will
indicate the instability (or stability) of the system.
\end{enumerate}

In all our numerical calculations we have chosen values for
the speed of sound and the fluid velocity close to unity. Moreover,
we set \mbox{$vc^2=1$} and choose units such that
\mbox{$\xi\,c = 1$}. The typical values of the velocity of sound in
BECs range between 1mm/s--10mm/s, while the healing length lies
between $10^{-3}$mm -- $10^{-4}$mm. In consequence, our numerical
results can be translated to realistic physical numbers by using
nanometres and microseconds as natural units. For example, the
typical lifetime for the development of an instability with
Im$(\omega) \simeq 0.1$ would be about $10$ microseconds.
We have checked that our results do not depend on
the particular values chosen for the velocities of the system.

\subsection{Black hole configurations}
\label{S:bh}

\begin{turnpage}
\begin{figure*}[htbp]\centering

\begin{tabular}[t]{l p{20pt} c p{40pt} c}

&& 

{\bf Black hole}

&&

{\bf White hole}

\\

&&
{\small $c_\text{super}=0.7$, $c_\text{sub}=1.8$.}

&&

{\small $c_\text{sub}=1.8$, $c_\text{super}=0.7$.}

\\

\begin{tabular}{c}
\vspace{-200pt}
\\
(a)
\\
\vspace{60pt}
\\
(b)
\\
\end{tabular}
&&
\includegraphics[height=.35\textwidth,clip]{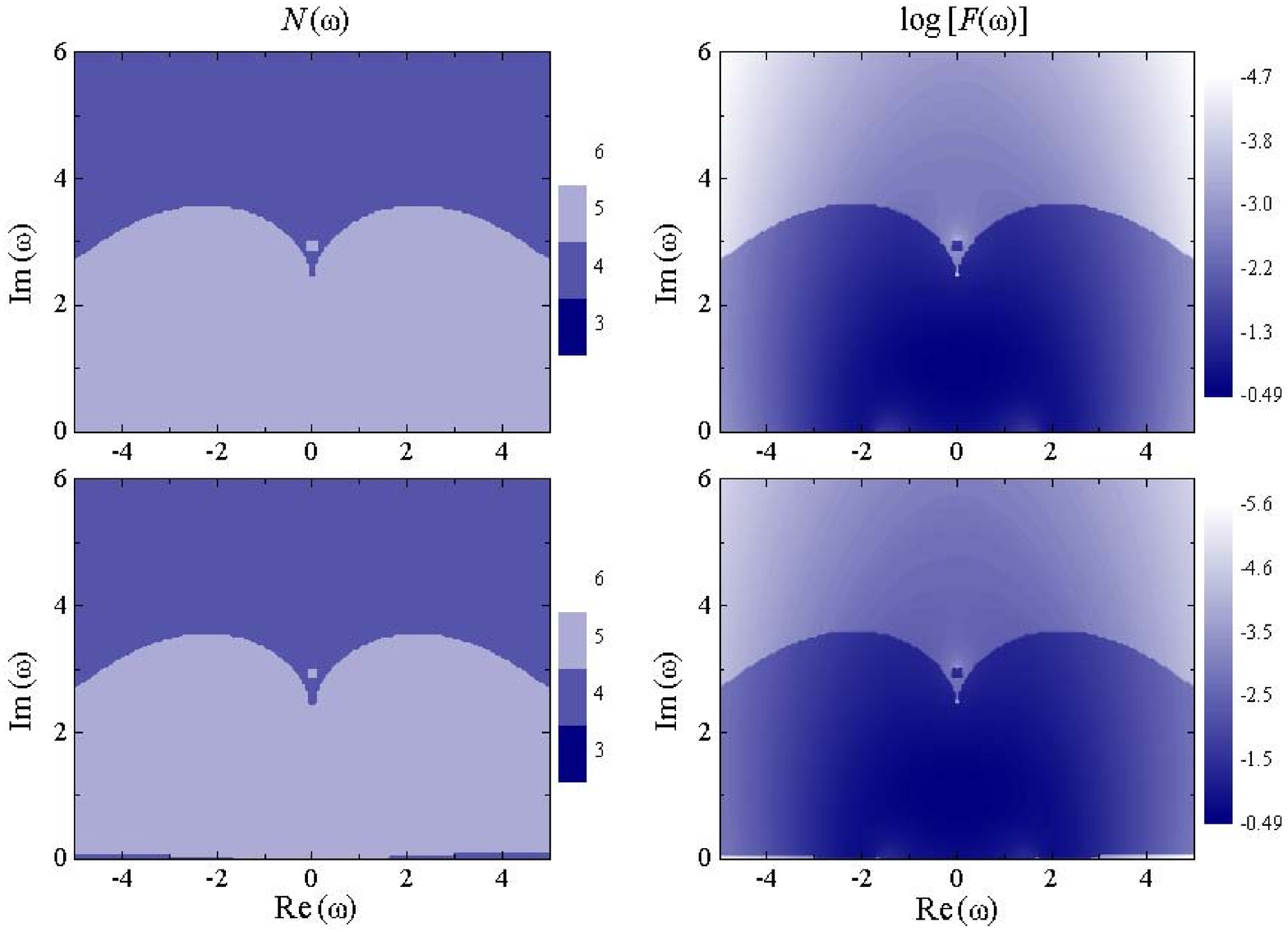}

&&

\includegraphics[height=.35\textwidth,clip]{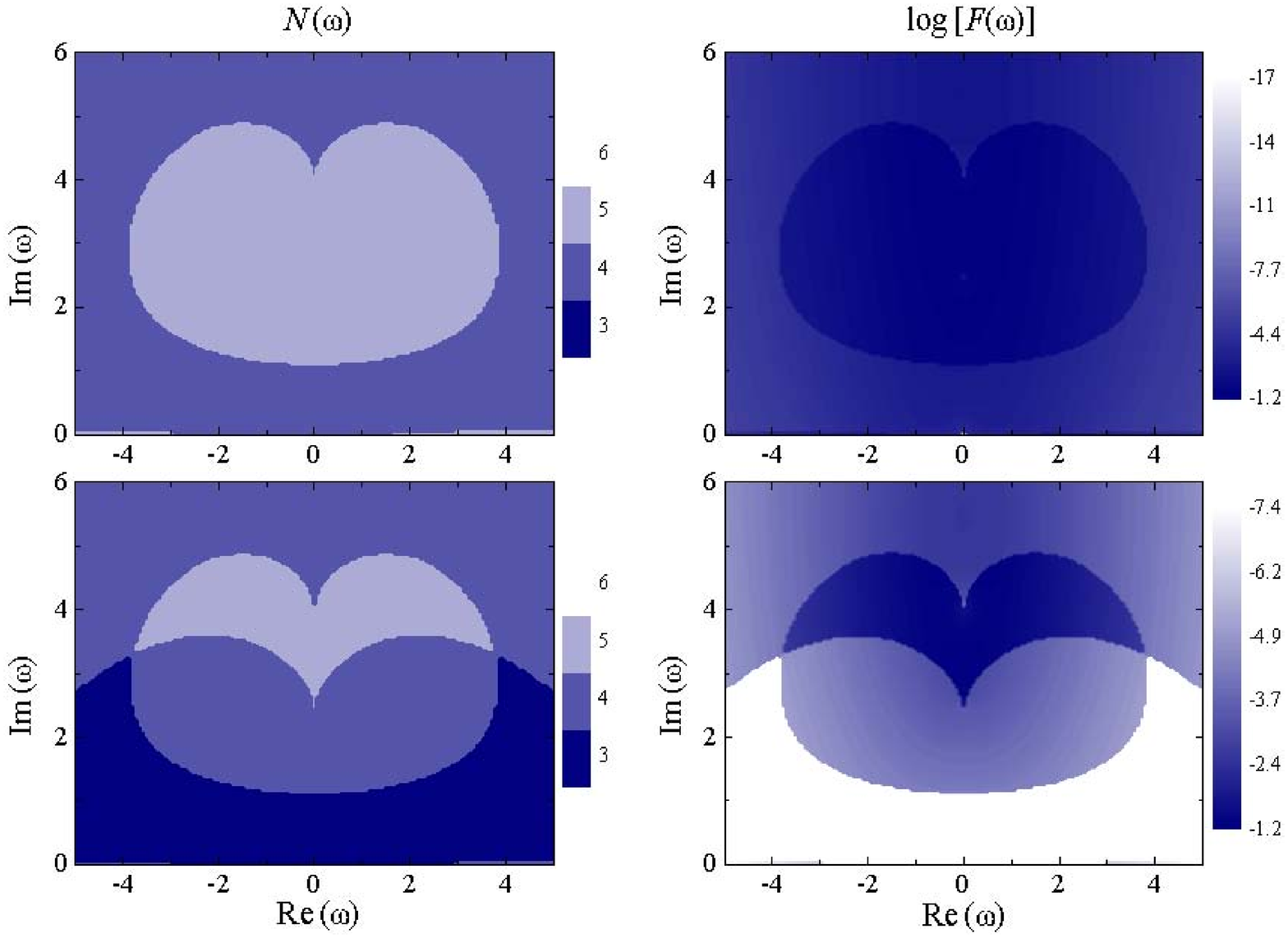}

\\ 

\vspace{0pt}

\\

&&
{\bf Accelerating subsonic flow}

&&

{\bf Decelerating subsonic flow}

\\

&&
{\small $c_\text{sub1}=1.8$, $c_\text{sub2}=1.9$.}

&&

{\small $c_\text{sub1}=1.9$, $c_\text{sub2}=1.8$.}

\\

\begin{tabular}{c}
\vspace{-200pt}
\\
(a)
\\
\vspace{60pt}
\\
(b)
\\
\end{tabular}

&&
\includegraphics[height=.35\textwidth,clip]{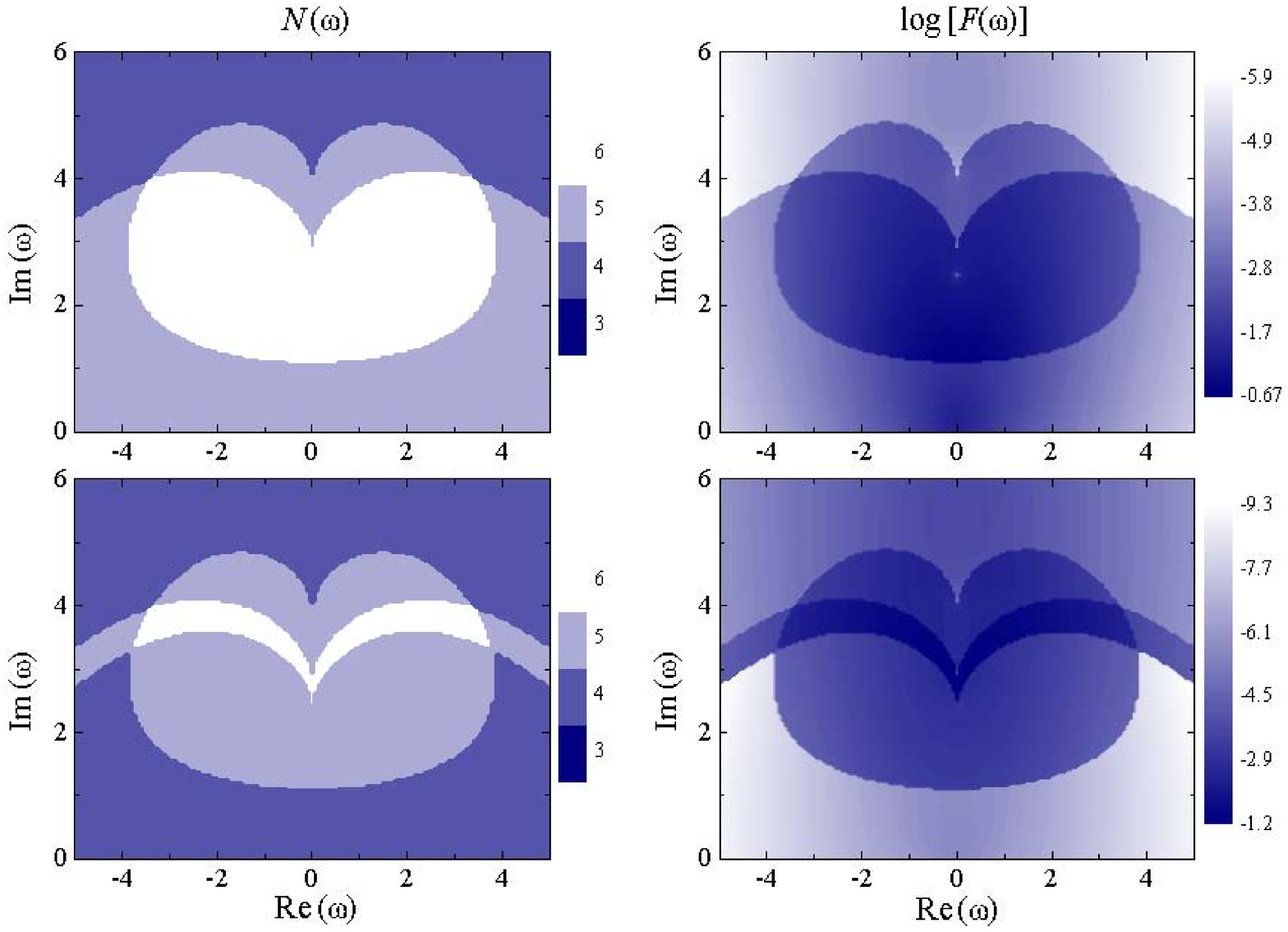}

&&
\includegraphics[height=.35\textwidth,clip]{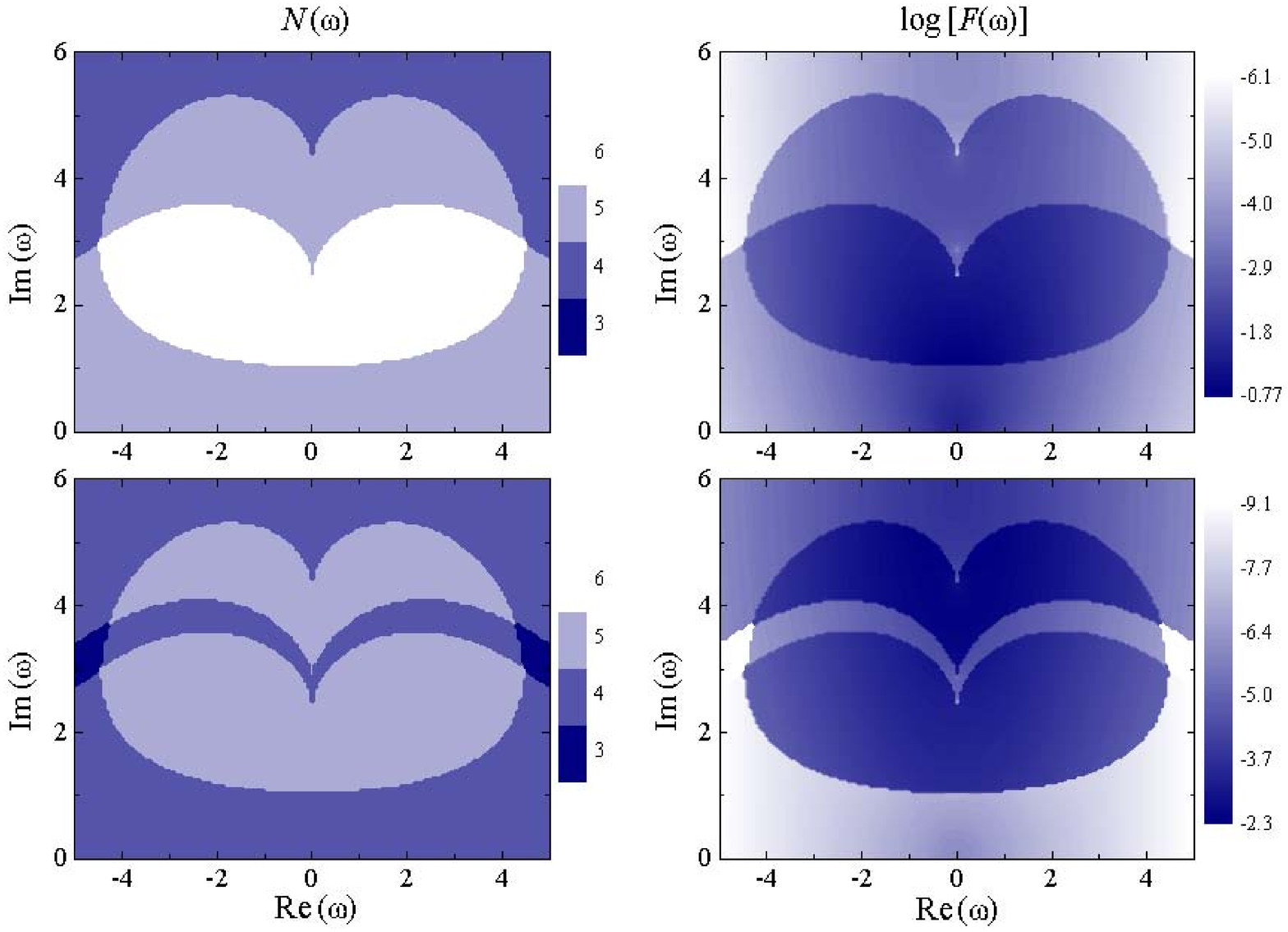}
\\ \end{tabular} 
\caption{Stability analysis under outgoing
boundary conditions for profiles with one discontinuity. Represented
is the relevant portion of the upper-half complex frequency
plane. From top to bottom and left to right: black hole and white hole
configurations, accelerating and decelerating subsonic flows (the
speed of sound $c$ is indicated for each region and the velocity $v$ is
then obtained from the constrain $vc^2=1$; $c>1$ corresponds to
a subsonic region, $c<1$ to a supersonic region; in addition, we use
$\xi \, c=1$ in all our calculations).  The lhs pictures
represent the number $N$ of forbidden modes in the asymptotic
regions. The rhs pictures represent the function $F(\omega)$ (to
enhance the contrast, we have drawn the logarithm), and white points
or regions, where $F(\omega)=0$, represent instabilities. In the upper
pictures [case (a)], convergence has been imposed in both asymptotic
regions. In the lower pictures [case (b)], convergence has been
imposed only in the upstream asymptotic region.  It is seen that black
hole configurations are stable in both case (a) and (b), as are
accelerating subsonic flows. White hole configurations are stable in
case (a), but develop a huge continuous region of instabilities in
case (b). Only a small strip of instabilities subsists in the
decelerating subsonic flow, indicating that the major part of this
unstable region is a genuine consequence of the existence of the white
hole horizon. Note that continuous regions of instability correspond
to $N<4$.}
\label{F:1steps} 
\end{figure*} 
\end{turnpage}

\begin{figure}[tbp]
\includegraphics[width=.3\textwidth,clip]{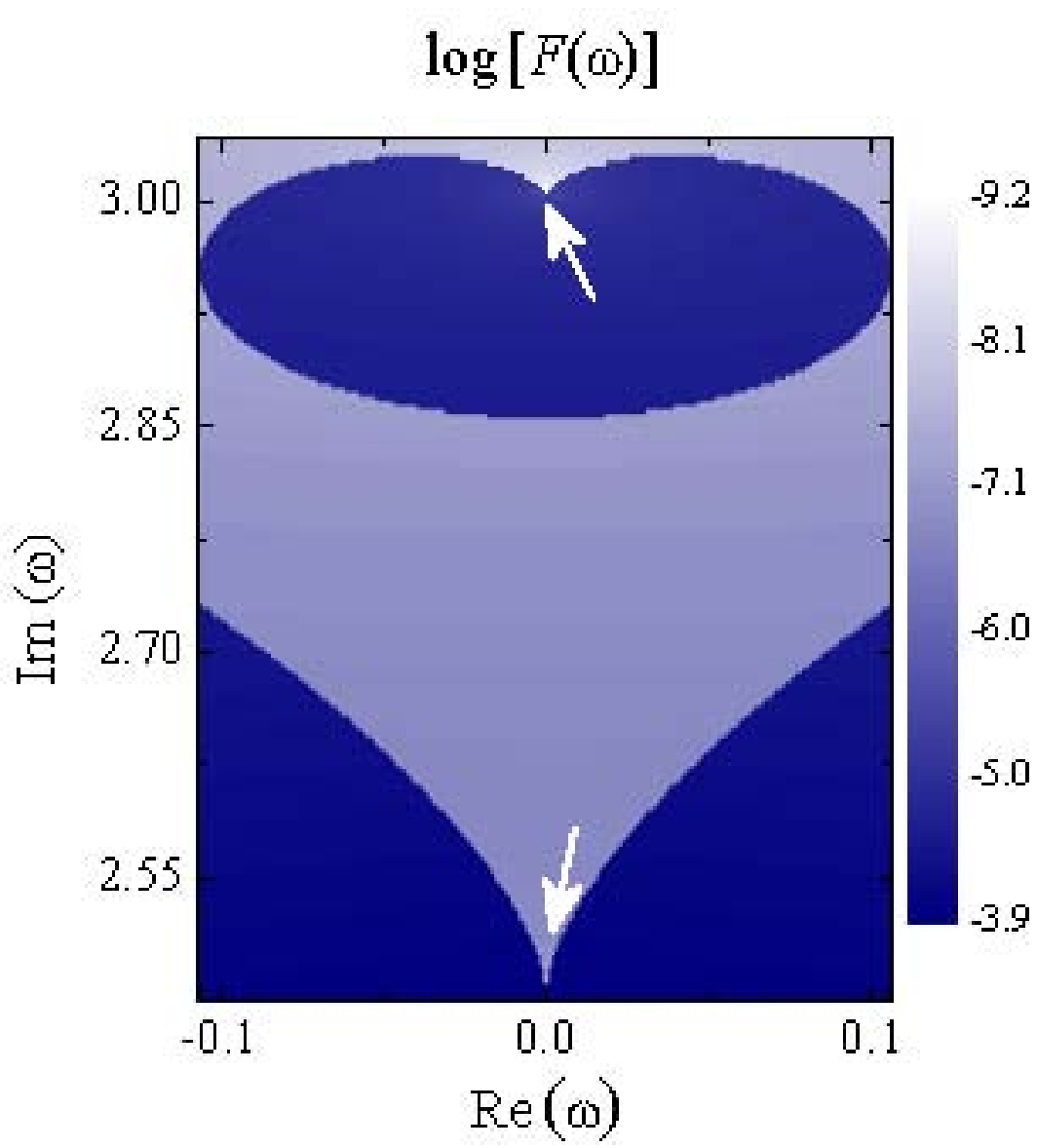}
\bigskip
\caption{Two special zeros of the function $F(\omega)$ appear in the stability 
analysis of a black hole configuration (this plot is a zoom of
the corresponding plot in Fig.~\ref{F:1steps}). They are
located at the boundary between regions with different number $N$ of
prohibited modes. These points do not seem to represent real 
instabilities of the system (see appendix~\ref{A:Zeros_Ceretes}).}
\label{F:Zoom}
\end{figure}

Consider a flow accelerating from a subsonic regime on the rhs to a
supersonic regime on the lhs, see Figure~\ref{F:v-super-sub}. For rhs
observers this configuration possesses a black hole horizon. For such
configurations with a single black hole-like horizon, when requiring
convergence in both asymptotic regions [case (a)], there are no zeros
(see Figure~\ref{F:1steps}), except for two isolated points on the
imaginary axis (see Figure~\ref{F:Zoom}, which is a zoom of the
relevant area in Figure~\ref{F:1steps}). (We always check the
existence of a zero by zooming in on the area around its location up 
to the numerical
resolution of our program.) These points are of a very special
nature. They are located at the boundary between regions with
different number $N$ of forbidden modes in the asymptotic regions. The
zeros that we will find for other configurations are of a totally
different nature: they are sharp vanishing local minima of
$F(\omega)$ living well inside an area with a constant value of $N$
($N=4$ to be precise). We discuss the meaning of these
special points in Appendix \ref{A:Zeros_Ceretes}. For now, let it
suffice to mention that points of this kind are always present in any
flow, independently of whether it reaches supersonic regimes or
not. Hence it seems that they do not correspond to real physical
instabilities, since otherwise any type of flow would appear to be
unstable. Accordingly, in the following, we will not
take these points into consideration. When we assert that a figure is
devoid of instabilities, we will mean that the function $F(\omega)$
has no zeros except for the special ones just mentioned.

Figure~\ref{F:1steps} also shows that the system remains
stable even when eliminating the condition of convergence at the
lhs [case (b)].

To sum up, configurations possessing a (single) black hole
horizon are stable under the general boundary conditions that we
have described, i.e. outgoing in both asymptotic regions and
convergent in the upstream asymptotic region, independently of
whether convergence is also fulfilled in the downstream asymptotic
region or not.

\subsection{White hole configurations}
\label{S:wh}

Let us now consider flows decelerating from a supersonic regime (rhs)
to a subsonic one (lhs). From the point of view of lhs observers, the
geometric configuration possesses a white hole horizon. In GR, a white
hole corresponds to the time reversal of a black hole.  Therefore,
unstable modes of a white hole configuration would correspond to
stable (quasinormal) modes of the black hole. However, when modified
dispersion relations are present, the precise definition of
quasinormal modes cannot be based only upon the outgoing character of
the modes, but their divergent or convergent character also has to be
taken into account.  Having in mind that in the acoustic approximation
(proper Lorentzian behaviour) the outgoing character of a quasinormal
mode implies that this mode diverges at the boundaries at infinity, it
is reasonable to impose divergence as an additional defining
requirement (apart from being outgoing) for a quasinormal mode in the
presence of modified dispersion relations. Using this definition, we
have checked (by analyzing the lower-half complex $\omega$ plane),
that black hole configurations do not show any quasinormal (stable)
eigenfrequency. Thus, we can conclude that {\em one-dimensional} white
holes are stable. We emphasize here the word one-dimensional because
we do not expect this situation to remain true in higher
dimensions. We know for example that standard GR black holes in 3+1
dimensions possess quasinormal modes. We expect these quasinormal
modes to subsist when taking into account departures from the acoustic
(Lorentzian) dispersion relation; we only expect them to acquire
modified eigenfrequencies. These quasinormal modes would then identify
instabilities of the corresponding white hole configuration. We leave
the analysis of the quasinormal modes in different analogue
gravitational configurations in BECs for future work, since this
analysis has its own subtleties.

The boundary conditions appropriate for the analysis of
white hole-like configurations correspond to only having ingoing waves
(due to time reversal) at the boundaries. But from the point of view
of acoustic models in a laboratory, the analysis
of the intrinsic stability of the flow (under the outgoing
boundary conditions described above) is also interesting. This analysis also has particular relevance with regard to configurations with two horizons (see below).

In Fig.~\ref{F:1steps} we see that under outgoing boundary conditions the
flow is stable when convergence is required in both asymptotic
regions [case (a)], but exhibits a continuous region of
instabilities at low frequencies when convergence is fulfilled only at
the rhs [case (b)]. Indeed, in this continuous region $N=3$,
in other words the algebraic system
$\widetilde\Lambda_{ij}A_{ij}=0$ is underdetermined and any frequency
is automatically an eigenfrequency.

When looking at the case of a completely subsonic flow suffering a
deceleration (see Fig.~\ref{F:1steps}), we find something similar. The
system is clearly stable when convergence is imposed at the lhs,
i.e. in case (a). Without convergence at the lhs, case (b), there is a
continuous strip of instabilities which corresponds, as in the
white hole case, to a region where $N=3$. However, this region
is localized at relatively high frequencies and so disconnected from
$\omega \sim 0$. We can say that part of the continuous region of
instabilities found in the white hole configuration has its origin merely in the
deceleration of the flow (giving rise to this high frequency strip).
But there is still a complete region of instabilities
that is genuine of the existence of a white hole horizon. In fact, by
decreasing the healing length parameter $\xi$, the strip moves up to
higher and higher frequencies, becoming less and less important as one
approches the acoustic limit. However, the continuous region of
instabilities associable with the horizon does not change its
character in this process.

We can therefore conclude the following with regard to
decelerating configurations. When convergence is fulfilled
downstream, the configuration is stable, regardless of whether it
contains a white hole horizon or not. When this convergence
condition is dropped, there is a tendency to destabilization. In the
presence of a white hole horizon, the configuration actually becomes
dramatically unstable, since there is a huge continuous region of
instabilities, and even perturbations with arbitrarily
small frequencies destabilize the configuration. In the absence of such a
horizon, only a small high-frequency part of this unstable region subsists.

\subsection{Black hole--white hole configurations}
\label{S:bh-wh}

\begin{figure}[tbp]
\includegraphics[width=.4\textwidth]{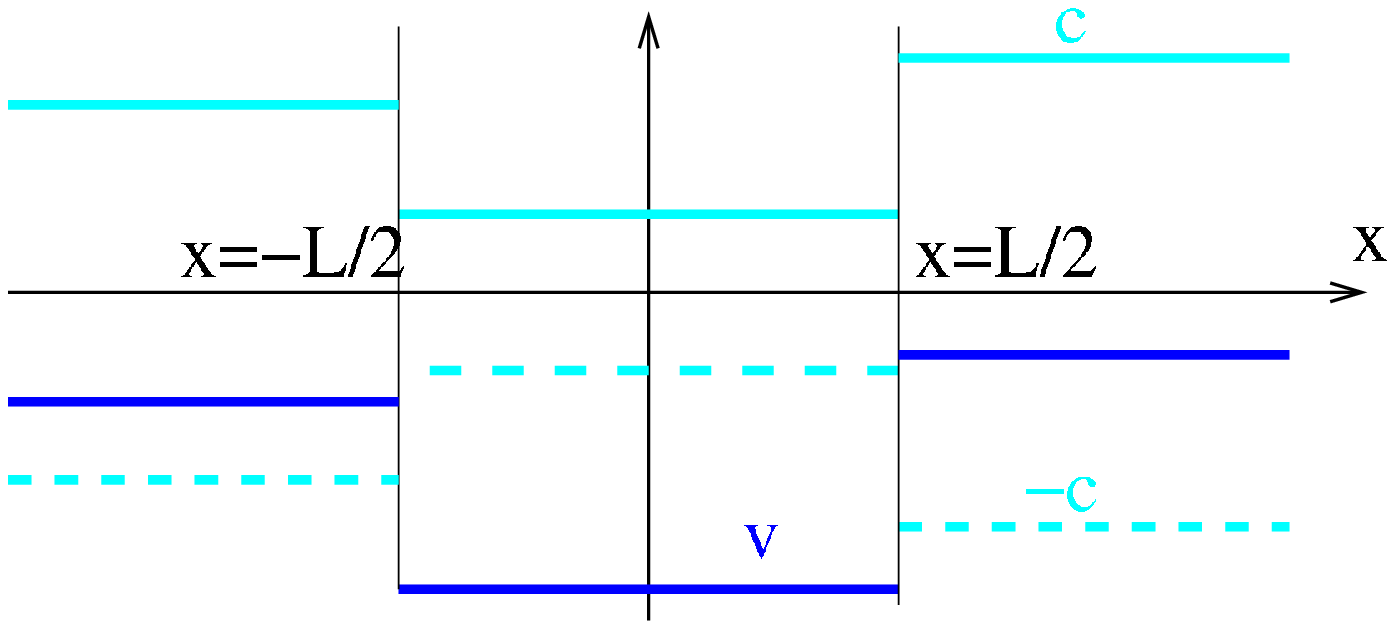}
\caption{Flow and sound velocity profiles with step-like discontinuities
simulating a black hole--white hole configuration.}
\label{F:v-sub-super-sub}
\end{figure}

Consider flows passing from being subsonic to supersonic and then back
to subsonic (Figure~\ref{F:v-sub-super-sub}). The numerical algorithm
we have followed to deal with this problem is equivalent to the one
presented above, but with a larger set of equations. In this case we
have $12$ arbitrary constants $A_j$, which have to satisfy $8+N$
equations: 4 matching conditions at each discontinuity and $N (0-8)$
additional conditions of the form $A_j=0$, corresponding to modes that
do not fulfill the boundary conditions in a particular asymptotic
region.

\begin{turnpage}
\begin{figure*}[tbp]\centering

\begin{tabular}[t]{l p{20pt} c p{40pt} c}

&& 

{\bf Black hole--white hole (globally accelerating)} 

&&

{\bf Black hole--white hole (globally decelerating)}

\\

&&

{\small $c_\text{sub-lhs}=1.8$, $c_\text{super}=0.7$, $c_\text{sub-rhs}=1.9$.}

&&

{\small $c_\text{sub-lhs}=1.9$, $c_\text{super}=0.7$, $c_\text{sub-rhs}=1.8$.}

\\

\begin{tabular}{c}
\vspace{-200pt}
\\
(a)
\\
\vspace{60pt}
\\
(b)
\\
\end{tabular}

&&

\includegraphics[height=.35\textwidth,clip]{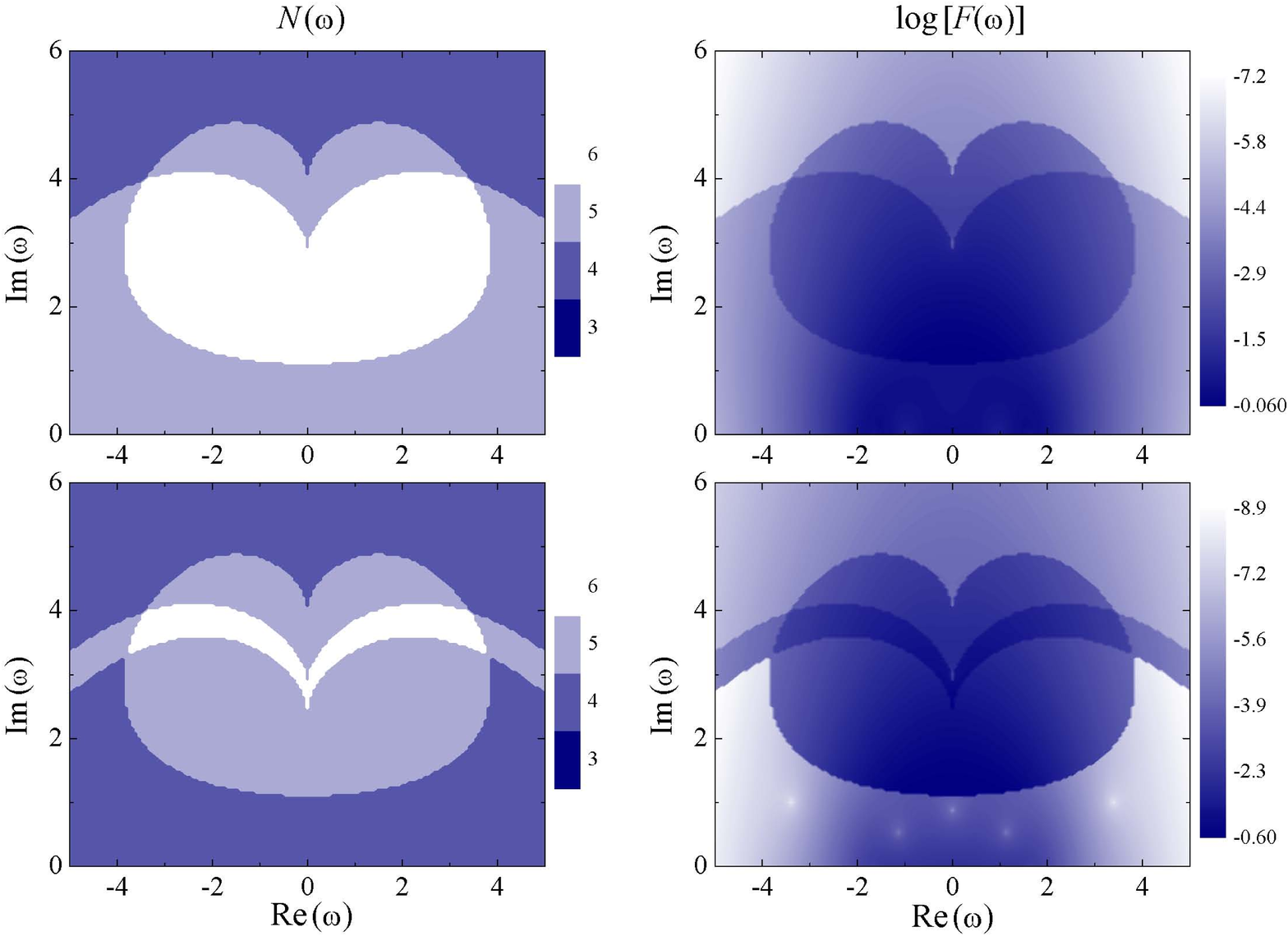}

&&

\includegraphics[height=.35\textwidth,clip]{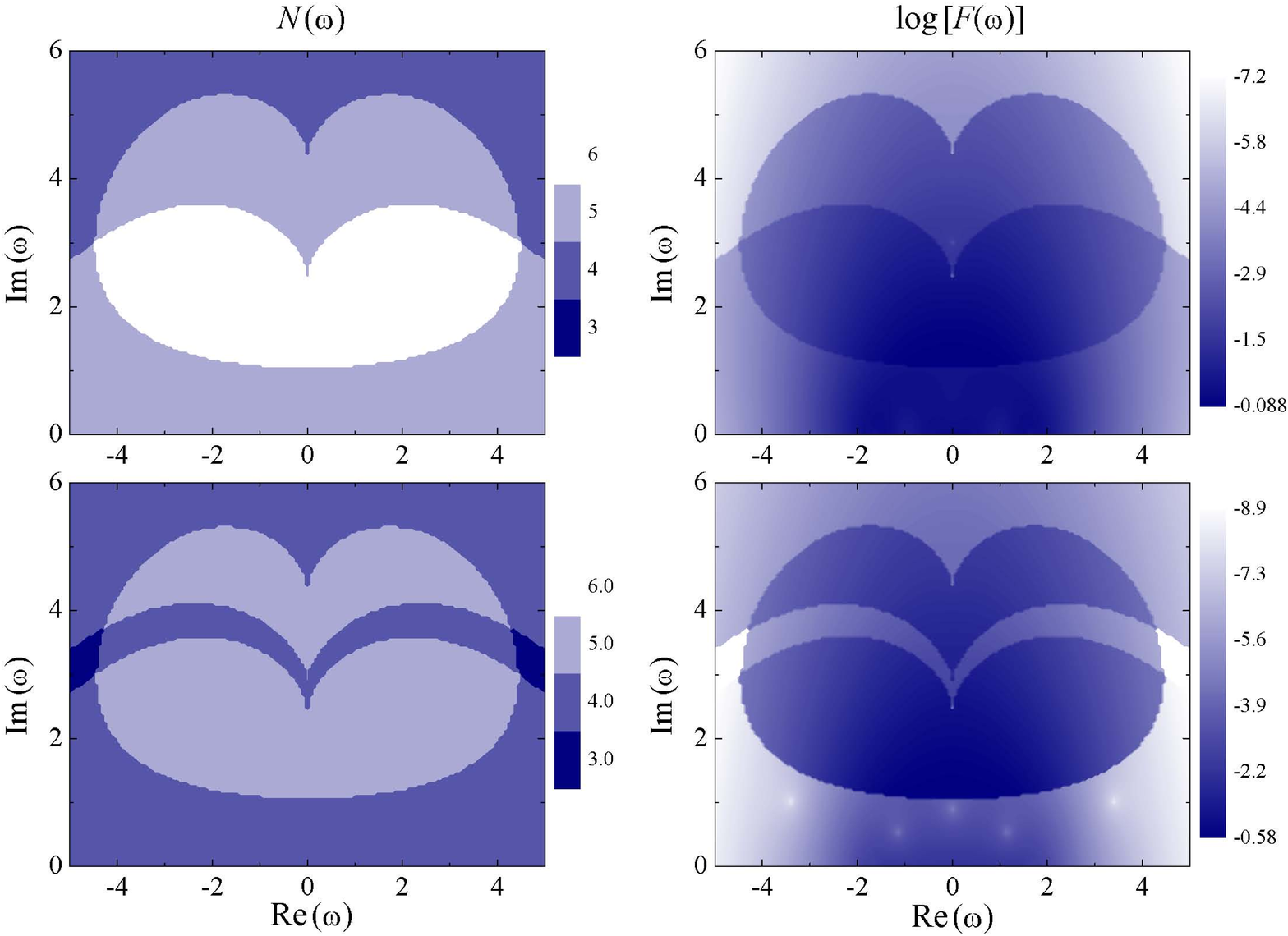}

\\

\vspace{0pt}

\\

&&

{\bf Globally accelerating subsonic flow}

&&

{\bf Globally decelerating subsonic flow}

\\

&&

{\small $c_\text{sub-lhs}=1.7$, $c_\text{sub}=1.8$, $c_\text{sub-rhs}=1.9$.}

&&

{\small $c_\text{sub-lhs}=1.9$, $c_\text{sub}=1.8$, $c_\text{sub-rhs}=1.7$.}

\\

\begin{tabular}{c}
\vspace{-200pt}
\\
(a)
\\
\vspace{60pt}
\\
(b)
\\
\end{tabular}
&&

\includegraphics[height=.35\textwidth,clip]{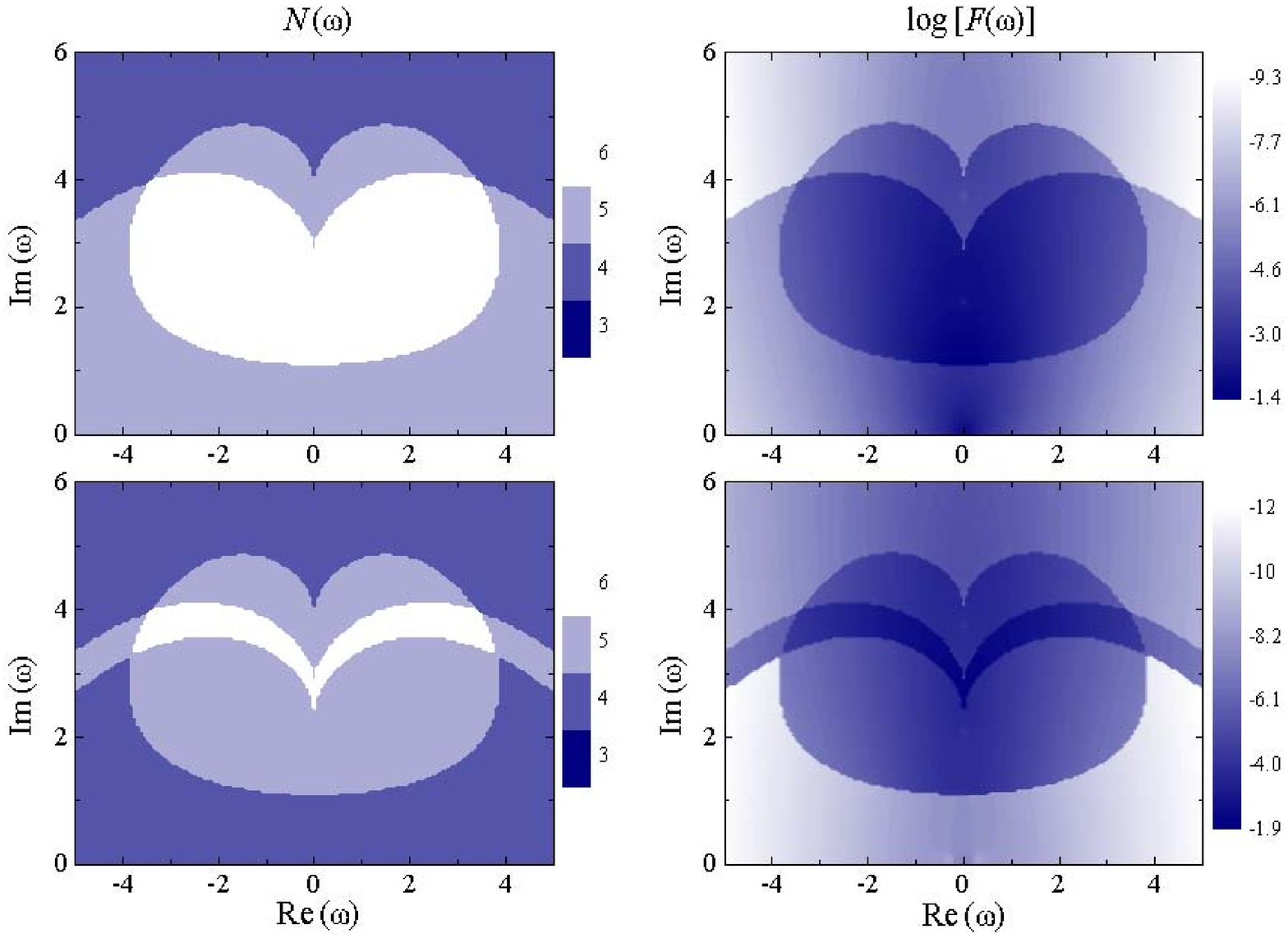}

&&
\includegraphics[height=.35\textwidth,clip]{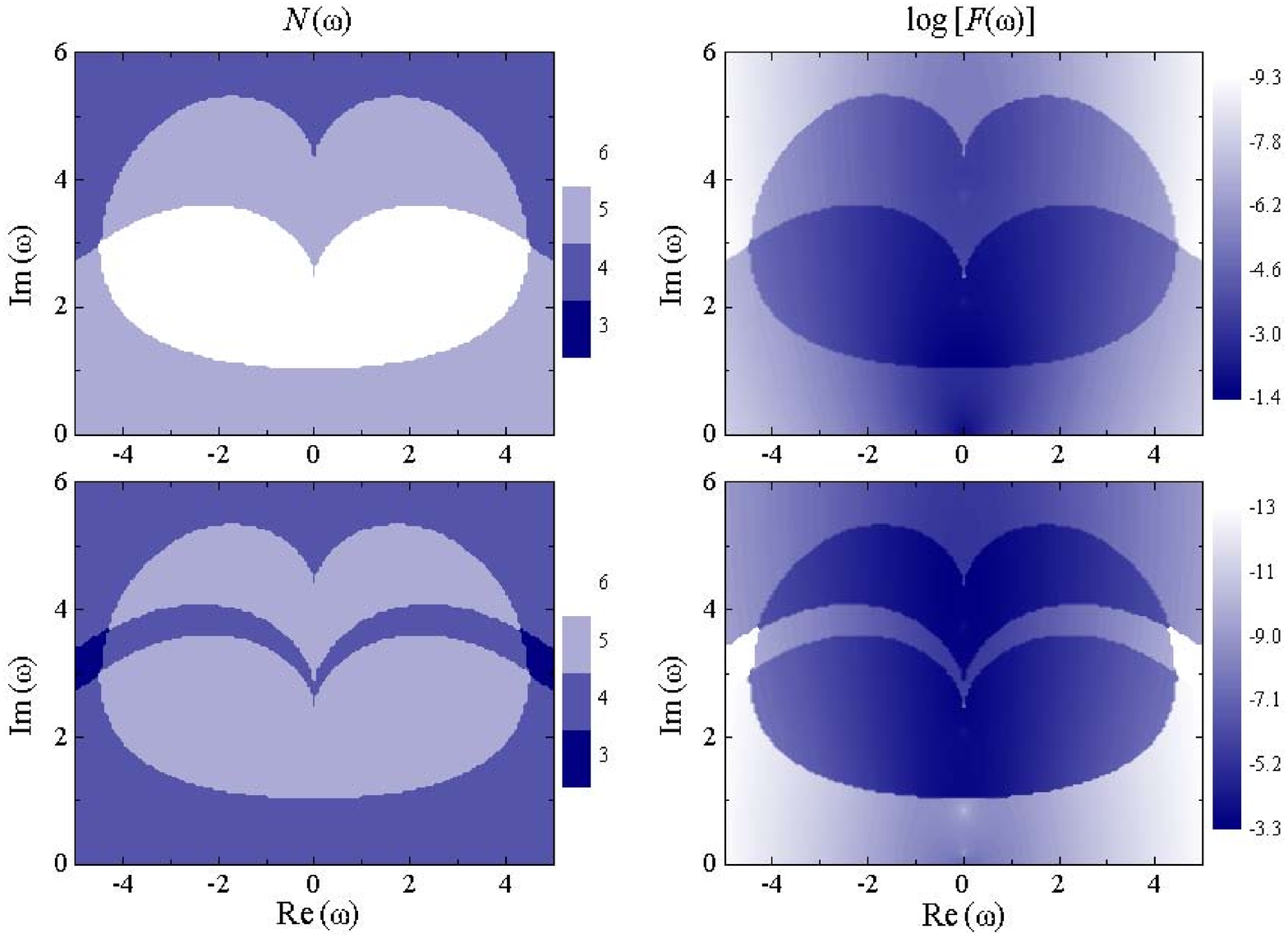}
\\ \end{tabular} 
\caption{Stability analysis for profiles with two
discontinuities. From top to bottom and left to right: globally
accelerating and decelerating black hole--white hole configurations,
globally accelerating and decelerating subsonic configurations.
In all these plots we have used $L=2.5$ as the size of the intermediate 
region (see also caption under Fig.~\ref{F:1steps}). When convergence is 
imposed in both asymptotic regions [case (a)], all the configurations are
stable. When convergence is only imposed upstream [case (b)], the
configurations with sonic horizons present a discrete set of
instabilities at low frequencies, while the decelerating
configurations show a small continuous unstable strip at high
frequencies. The decelerating configuration with sonic horizons
combines both types of instabilities.}  
\label{F:2steps} 
\end{figure*}
\end{turnpage}

When convergence is imposed at the lhs, we do not find any
instabilities, regardless of whether the fluid is globally
accelerating or decelerating [the final lhs fluid velocity is larger
or smaller than the initial rhs one respectively, see
Figure~\ref{F:2steps} cases (a)].  Also when replacing the
intermediate supersonic region by a subsonic one, thereby removing the
acoustic horizons, the fluid is stable, independently of whether it is
globally accelerating or decelerating.

When dropping the convergence condition at the lhs the situation
changes completely. When the intermediate region is supersonic,
i.e. in a black hole--white hole configuration, a discrete set of
instabilities appears at low frequencies [Fig.~\ref{F:2steps} cases
(b)]. It is worth mentioning that, when carefully looking at plots of
type (a)-cases, we observe some traces of these zeros in the form of
local minima which can be understood as particularly soft regions.
These regions, although very close to zero in some situations, never
give rise to real zeros, as we have carefully checked by zooming
in. Notice that these local minima appear in regions with $N=5$ where
a zero would mean a double degeneracy within the row vectors in the
corresponding matrix $\tilde \Lambda_{ij}$. When the fluid is globally
decelerating, additionally there is a continuous region of
instabilities at higher frequencies. Indeed, in this region, as in the
case of the white hole configuration, $N<4$, and so every frequency in
this region automatically represents an instability. When the
intermediate region is subsonic, the discrete set of local minima at
low frequencies disappears, but the continuous strip of instabilities
at higher frequencies persists in the case of a globally decelerating
fluid.  The discrete set of instabilities is therefore a genuine
consequence of the existence of horizons.

\begin{figure}[tbp]\centering
\includegraphics[height=.75\textwidth,clip]{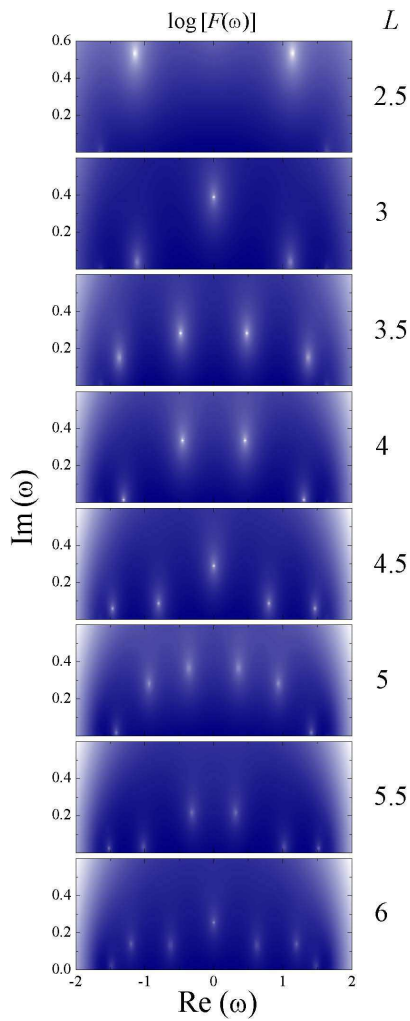}
\caption{The number of discrete instabilities in the black hole--white hole 
configuration increases with the size $L$ of the intermediate
supersonic region. In these plots we have used $c_{\rm sub-lhs}=1.8$,
$c_{\rm super}=0.7$, $c_{\rm sub-rhs}=1.9$ (with their corresponding 
$v=1/c^2$) and $\xi \,c=1$.}
\label{F:2steps_more}\end{figure}

The number of discrete zeros we find in the black hole--white hole
configuration increases with the size $L$ of the supersonic region
(see Fig. \ref{F:2steps_more}), while their $\text{Im}(\omega)$
decreases. This suggests that the region between the horizons acts as
a sort of well discretizing some of the instabilities found for
the white hole configurations. The larger the well, the larger the
amount of instabilities, but the longer-lived these instabilities.

To summarize, when requiring convergence in both asymptotic regions,
all the types of configurations with two discontinuities that we have
discussed are stable. When not requiring convergence at the lhs,
discretized instabilities appear associated with the presence of
horizons.

\subsection{Black hole configurations with modified boundary conditions}
\label{S:sink-bh}

\begin{figure}[tp]\centering
\begin{minipage}{.45\textwidth}\centering
\includegraphics[width=.7\textwidth,clip]{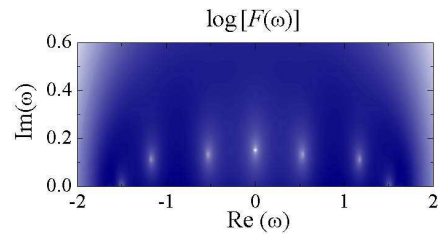}
\end{minipage}
\caption{A discrete set of instabilities appears in a black hole 
configuration when the lhs asymptotic region representing the
singularity is replaced by a wall or sink. In this plot we have used
$c_{\rm sub}=1.9$ and $c_{\rm super}=0.7$ (with their corresponding
$v=1/c^2$); in addition we have taken $L=6$ as the size of the, now
finite, internal region.}
\label{F:Sink}
\end{figure}

We have seen in section~\ref{S:bh} that configurations with a single
black hole horizon do not possess instabilities in any
situation. However, as we have just discussed, when they are continued
into a white hole configuration, some instabilities can show up. Here,
we would like to point out that the same happens if instead of
extending the black hole configuration we introduce a wall (or sink)
at a finite distance inside the supersonic region, described
by other boundary conditions than the ones we have considered so far.
For example, by replacing the lhs boundary conditions by $\theta|_{x=-L}=0$,
we obtain Fig.~\ref{F:Sink}. We can perfectly see how a set of
discrete unstable modes appears.

\section{Discussion and conclusions}
\label{S:discussion}

Let us start by discussing the stability of configurations with a
single black hole-like horizon in analogue systems that incorporate
superluminal dispersion relations. We have seen that by requiring
purely outgoing and convergent boundary conditions in both asymptotic
regions, these configurations do not show any signs of instability.
The same applies when dropping the convergence condition
downstream (i.e. on the lhs). This seems to contradict the results
in Ref.~\cite{sonicBH2}. There, the existence of a future (spacelike)
singularity inside the black hole, from which no information is
allowed to escape, was implemented by introducing a sink in the
supersonic region at a finite distance from the horizon. Then, it was
found that there were discrete instabilities in the system. However,
these instabilities correspond to the following particular set of
boundary conditions: i) At the asymptotic region, only convergent
boundary conditions were imposed, without any condition about the
direction of propagation (in- or outgoing) of the perturbations;
\mbox{ii) At} the sink, two types of boundary conditions were
required, specifically designed for dealing with symmetric and
anti-symmetric configurations. In our language these boundary conditions
correspond to $\{ \theta'|_{\rm sink} = 0, n'|_{\rm sink}=0 \}$ and $\{
\theta|_{\rm sink} = 0, n|_{\rm sink}=0 \}$ respectively. In comparing
this result with ours we have checked two important facts. On the one
hand, their unstable modes have ingoing contributions at the
asymptotic region. On the other hand, the boundary conditions at the
sink are such that they combine outgoing and ingoing contributions --
their sink implementation makes waves reaching the sink bounce back
towards the horizon. These two facts are responsible for the unstable
behaviour of these black hole-like configurations. If no energy is
introduced into the system from the asymptotic region (in other words,
if only outgoing perturbations are allowed) and moreover any bouncing
at the sink is eliminated, then these configurations
are stable. This is in agreement with the result found in
Ref.~\cite{leonhardt}.

In the case of configurations with a single white hole horizon, we
have seen that with outgoing and convergent boundary conditions in
both asymptotic regions, there are no instabilities in the
system. However, when eliminating the convergence condition in the
downstream asymptotic region, one finds a continuous region of
instabilities surrounding $\omega=0$. Thus, we see that these white
hole configurations are stable only when the boundary conditions are
sufficiently restrictive.

When analyzing configurations connecting two different subsonic
regions, we have also seen that, again, when convergence is required
at the lhs, they are stable. But when this convergence condition is
relaxed, globally decelerating configurations tend to become unstable,
whereas globally accelerating ones remain stable. The instabilities of
these decelerating configurations without horizons (i.e. purely
subsonic ones) show up, however, in a small strip at high
frequencies. In contrast, white hole configurations present
instabilities for a wide range of frequencies, starting from
arbitrarily small values. This points out that the presence of a white
hole horizon drastically stimulates the instability of the configuration.

With regard to the black hole--white hole configurations, we have seen that,
as before, with outgoing and convergent boundary conditions, they are
stable. However, when relaxing the convergence condition downstream,
they develop a discrete set of unstable modes.

In the analysis of the black hole laser instability in Ref.~\cite{jacobson},
the authors found that these black hole--white hole configurations
were intrinsically unstable. However, they did not analyze what
happens to the modes at the lhs infinity. Our analysis shows that by
restricting the possible behaviour of the modes in the downstream
asymptotic region, one can eliminate the unstable behaviour of
the black hole laser. This is in agreement with the results
in Refs.~\cite{sonicBH1,sonicBH2}. There, the instabilities can in some
cases be removed by requiring periodicity, that is, by imposing
additional boundary conditions to the modes.

To sum up, we have shown the high sensibility of the stability not
only on the type of configuration (the presence of a single horizon or
of two horizons, the accelerating or decelerating character of the
fluid), but particularly on the boundary conditions. With
outgoing boundary conditions, when requiring convergence at the
downstream asymptotic region, both black hole and white hole
configurations are stable (and also the combination of both into a
black hole--white hole configuration). When relaxing this convergence
condition at the lhs, configurations with a single black hole horizon
remain stable, whereas white hole and black hole--white hole
configurations develop instabilities not present in (subsonic) flows 
without horizons.

\section*{Acknowledgements}

C.B. has been funded by the Spanish MEC under project
FIS2005-05736-C03-01 with a partial FEDER contribution. G.J. was
supported by CSIC grants I3P-BGP2004 and I3P-BPD2005 of the I3P
programme, cofinanced by the European Social Fund,
and by the Spanish MEC under project FIS2005-05736-C03-02.
L.G. was supported by the Spanish MEC under the same project and FIS2004-01912.

\appendix

\section{Zeros at the boundaries of the regions in $N(\omega)$
\label{A:Zeros_Ceretes}}

Given an $\omega$, one can find its four associated $k$ roots,
$\{k_j\}$. If instead of $\omega$ one takes $\tilde \omega= -\omega^*$,
it can be seen that the new roots $\{\tilde k_j\}$ are just
$\{-k_j^*\}$. For this reason the function $F(\omega)$ is
mirror symmetric with respect to the imaginary axis (this is seen in
all our figures). Now, when $\omega$ is pure imaginary ($\omega=
-\omega^*$), the set $\{k_j\}$ has to be equal to the
set $\{-k_j^*\}$. There are three posibilities. Either all four roots
are pure imaginary, two are imaginary and the other two complex
satisfying $k_j=-k_l^*$, with $j \neq l$, or there are two pairs of
complex roots satisfying $k_j=-k_l^*$. When moving through the
imaginary $\omega$ axis, there are points at which there is a
transition from one of these possibilities to another. At any
transition point there has to be a pair of imaginary roots with equal
value. Defining $\omega'' \equiv {\rm Im}(\omega)$ and $\kappa \equiv
-ik$, the dispersion equation~(\ref{dispersion}) can be written as
\begin{align}
(\omega''- v\kappa)^2
-\left(c^2 - {1 \over 4} c^2 \xi^2 \kappa^2\right)\kappa^2 =0.
\label{dispersion-modified}
\end{align}
This is a fourth order polynomial in $\kappa$ with real coefficients.
If this polynomial has two equal real roots then we know that the derivative
with respect to $\kappa$ of the polynomial has to be zero at that point.
It is not difficult to see that this also implies that the derivative
with respect to $\kappa$ of the function
\begin{align}
(w- v\kappa)\mp
\left(c^2 - {1 \over 4} c^2 \xi^2 \kappa^2\right)^{1/2}\kappa
\end{align}
has to be zero at this same point. But this derivative coincides with
the definition of the group velocity given in~(\ref{group-velocity}) (when
$\omega$ and $k$ are pure imaginary, $d\omega/dk$ is directly real).
Therefore, we conclude that at any transition point
on the imaginary $\omega$-axis we have degeneracy: at least two imaginary $k$
roots with equal value. At the same time, the group
velocity associated with them becomes zero. This is why these points
are located at the boundary between regions with a different number of
forbidden modes: these are places in which outgoing modes transform
into ingoing ones. The zero that appears in the function $F(\omega)$ at these
points is due to the degeneracy and does not tell us anything about
the existence or not of a real instability there. To know whether a real instability appears, one first has to find the actual four independent solutions of
equations~\eqref{GP_lin} at the point that led
to the degeneracy. Let us check under which circumstances
one can find a solution of the form
\begin{subequations}\label{plane-waves_app}\begin{align}
\widetilde n_1(x,t)&= A_1 \, x \,e^{i(k x - \omega) t},
 \\
\theta_1(x,t)&= B_1 \, x \,e^{i(k x - \omega) t}+ B_2 e^{i(k x - \omega) t}.
\end{align}\end{subequations}
For these expressions to be a solution of Eqs.~\eqref{GP_lin}, the following conditions
have to be satisfied:
\begin{widetext}
\begin{gather}
\begin{pmatrix}
i(\omega- vk)  && c^2 k^2 \\ \\
1 +{1 \over 4} \xi^2 k^2 && - i(\omega-vk)
\end{pmatrix}
\begin{pmatrix}
A_1 \\ \\
B_1
\end{pmatrix}
=0,
\\
\nonumber
\\
\begin{pmatrix}
i(\omega- vk)  && c^2 k^2 \\ \\
1 +{1 \over 4} \xi^2 k^2 && - i(\omega-vk)
\end{pmatrix}
\begin{pmatrix}
0 \\ \\
B_2
\end{pmatrix}
+
\begin{pmatrix}
-v && -2 i c^2 k \\ \\
-{1 \over 2} i \xi^2 k && v
\end{pmatrix}
\begin{pmatrix}
A_1\\ \\
B_1
\end{pmatrix}
=0.
\end{gather}
\end{widetext}%
{}From the first condition we obtain that the dispersion relation
\eqref{dispersion} has to be fulfilled. As a consequence we also find
that $B_1=A_1 (\omega -v k)/(ic^2 k^2)$. Now, from the second
condition we obtain
\begin{align}
\begin{pmatrix}
-c^2 k^2 \\ \\
 i(\omega-vk)
\end{pmatrix}
=
{A_1 \over B_2}
\begin{pmatrix}
-v && -2 i c^2 k \\ \\
-{1 \over 2} i \xi^2 k && v
\end{pmatrix}
\begin{pmatrix}
1\\ \\
 {(\omega -v k) \over ic^2 k^2}
\end{pmatrix}
\end{align}
This is a system of two equations from which, eliminating
$A_1/B_2$ and after some rearranging, we obtain:
\begin{align}
c^2 k +{1 \over 2}c^2 \xi^2 k^3 + v(\omega -vk)=0.
\end{align}
This is exactly the condition for a vanishing group
velocity~\eqref{group-velocity}. Therefore, when functions in the form
of plane waves do not lead to four linearly independent solutions, but
for example two are ``degenerate'', then we can use the previous
solution~\eqref{plane-waves_app} avoiding this degeneracy. Once we
have the actual four independent solutions of the problem, they have
to be matched with the four solutions in the other region (typically
these will have the form of plane waves, unless we are in a very
special situation in which degeneracy occurs in both regions at the
same time) and see whether there is a combination satisfying all the
boundary conditions.

Although we haven't made such a full detailed calculation, the fact
that this kind of situation occurs in any type of flow indicates that
it is safe to assume that they do not represent real instabilities, as
already mentioned.

\section{Bogoliubov representation}\label{A:app}

All calculations and numerical simulations presented in the text have
been performed independently by using the acoustic representation and
the Bogoliubov representation \cite{sonicBH1, sonicBH2} described in
this appendix.  We have found identical results with the two methods,
double checking in this way the absence of numerical artifacts.

Consider a one-dimensional setup with a potential $V_{\rm ext}$ that
produces a profile for the speed of sound of the form
\begin{equation}
c(x)= \left\{\begin{array}{ll} c_0, & x<0
\\
c_0 [1+(\sigma-1)x/\epsilon], \quad& 0<x<\epsilon
\\
\sigma c_0, & \epsilon<x
\end{array}\right. .
\end{equation}
We will assume $\sigma >1$ and a flow velocity in the inward
($x\to-\infty$) direction, i.e. a black hole-like configuration for a
rhs observer. The limit $\epsilon \rightarrow 0$ provides the same
profile we have discussed in the main text.

The condensate wave function can be written as the sum of a
stationary background state $\psi_0$ and a perturbation $\phi$
satisfying
\begin{equation}\label{eq:pertcond}
i\partial_t\phi=-\partial_x^2\phi/2+(c^2-v^2/2+\partial_x^2 c/2c)\phi + c^2e^{2i\int^x v}\phi^*,
\end{equation}
that will then be expanded into Bogoliubov modes
\begin{equation}
\phi = u_{\omega}(x) e^{-i\omega t}+ w^*_{\omega}(x) e^{i\omega^* t}.
\end{equation}
>From now on, we drop the subindex $\omega$, but remember that all
equations should be valid for every (complex) frequency $\omega$ separately.

The assumption of a small $\epsilon$ leads to a linear solution for
the modes $u$ and $w$ in the intermediate region
\mbox{$0<x<\epsilon$}. Together with the transition conditions at
$x=0$ and $x=\epsilon$, which substitute the singular character of
$\partial_x^2 c/c$ at those points, this leads to the following connection
formulas (in the limit \mbox{$\epsilon \rightarrow 0$}), see
Ref. \cite{sonicBH2}:
\begin{align}\label{uporc}
[u/c]=&0, & [\partial_x(cu)]=&0,
\\
[w/c]=&0, & [\partial_x(cw)]=&0,
\end{align}
where as before we have used the notation
\mbox{$[u]=u|_{x \to 0^+} -u|_{x \to 0^-}$.}

Let us write the density and phase in terms of the stationary values
plus perturbations:
\begin{equation}
n=n_0+\g^{-1}\widetilde n_1,\qquad \theta=\theta_0+\theta_1,
\end{equation}
and note that the matching condition for the background velocity
potential is $[\theta_0]=0.$

The relation between the acoustic and the Bogoliubov
representation of the perturbations can easily be obtained by noting
that, to first order,
\begin{equation}
\phi=\sqrt{m/\g} \, c(\widetilde n_1/2mc^2+i\theta_1)e^{i\theta_0},
\label{eq:psintheta}
\end{equation}
where we have used the fact that $c^2=\g n_0/m$. Note that
$\widetilde n_1$ and $\theta_1$ are real. Then, comparison of
Eq. \eqref{eq:psintheta} with the Bogoliubov mode expansion \mbox{$
\phi= u e^{-i\omega t}+w^* e^{i\omega^*t}$} yields
\begin{align}
u&=\sqrt{m/\g} \, c(\widetilde n_1/2mc^2+i
\theta_1)e^{i\theta_0},\\
w&=\sqrt{m/\g} \, c(\widetilde n_1/2mc^2-i \theta_1)e^{-i\theta_0}.
\end{align}
We then have
\begin{equation}
[u/c]\propto[(\widetilde n_1/2c^2+i
\theta_1)e^{i\theta_0}]\propto[\widetilde n_1/2c^2+i \theta_1]
\end{equation}
and likewise for $w$. Comparing with Eq. \eqref{uporc}, we obtain
\begin{gather}
[\widetilde n_1/c^2]\propto[u/c]+[w/c]=0,
\\
[\theta_1]\propto[u/c]-[w/c]=0,
\end{gather}
which correspond to two of the conditions in Eqs.~\eqref{matching}.

Taking into account that $[\theta_0]=[c^2v]=[\theta_1]=0$ and
that $\partial_x c_L=\partial_x c_R=0$, we can write
\begin{align}
[\partial_x(cu)]\propto& \; i[v\widetilde n_1]/2+[\partial_x\widetilde n_1]/2 +i[c^2\partial_x\theta_1],
\\
[\partial_x(cw)]\propto& -i[v\widetilde n_1]/2+[\partial_x\widetilde n_1]/2 -i[c^2\partial_x\theta_1].
\end{align}
In other words \mbox{$[\partial_x\widetilde n_1]=0$} and
\mbox{$[v\widetilde n_1/2+c^2\partial_x\theta_1]=0$}.
Because of the continuity equation $vc^2=\text{const}$, we have
\mbox{$[v\widetilde n_1]\propto[n_1/c^2]=0$},
so that the boundary conditions ultimately become
\begin{align}\label{eq:boundcond}
[\partial_x\widetilde n_1]&=0,& [c^2\partial_x\theta_1]&=0,
\end{align}
i.e. the other two conditions in Eqs.~\eqref{matching} with the corresponding simplifications.


\end{document}